\documentclass[%
 aip,
 amsmath,amssymb,
preprint,%
]{revtex4-1}

\usepackage{graphicx}
\usepackage{dcolumn}
\usepackage{bm}

\usepackage[utf8]{inputenc}
\usepackage[T1]{fontenc}
\usepackage{mathptmx}
\usepackage{etoolbox}

\makeatletter
\def\@email#1#2{%
 \endgroup
 \patchcmd{\titleblock@produce}
  {\frontmatter@RRAPformat}
  {\frontmatter@RRAPformat{\produce@RRAP{*#1\href{mailto:#2}{#2}}}\frontmatter@RRAPformat}
  {}{}
}%
\makeatother
\begin{document}

\preprint{}

\title[]{The map between symmetries and orbital rules to realize tunable band gap in quantum anomalous Hall effect material}
\author{Jiaohong Shu}
\affiliation{
School of Mathematics, Physics and Statistics, Shanghai University of Engineering Science, Shanghai 201620, People's Republic of China.
}%
\author{Xinxin Zhao*}%
 \email{bighunter@sues.edu.cn}
\affiliation{
School of Mathematics, Physics and Statistics, Shanghai University of Engineering Science, Shanghai 201620, People's Republic of China.
}%

\author{Weiqin Fan}
\affiliation{
School of Mathematics, Physics and Statistics, Shanghai University of Engineering Science, Shanghai 201620, People's Republic of China.
}%

\author{Yunjiu Cao}
\affiliation{
School of Mathematics, Physics and Statistics, Shanghai University of Engineering Science, Shanghai 201620, People's Republic of China.
}%
\author{Lili Wang}
\affiliation{
School of Mathematics, Physics and Statistics, Shanghai University of Engineering Science, Shanghai 201620, People's Republic of China.
}%

\author{Guanglong Chen}
\affiliation{
School of Mathematics, Physics and Statistics, Shanghai University of Engineering Science, Shanghai 201620, People's Republic of China.
}%

\author{Jianbao Wu}
\affiliation{
School of Mathematics, Physics and Statistics, Shanghai University of Engineering Science, Shanghai 201620, People's Republic of China.
}%

\author{Yiming Mi}
\affiliation{
School of Mathematics, Physics and Statistics, Shanghai University of Engineering Science, Shanghai 201620, People's Republic of China.
}%

\date{\today}

\begin{abstract}
We establish the map between symmetries and orbital rules to realize tunable band gap in quantum anomalous Hall effect material. The band gap is determined by the spin-orbit coupling between local orbitals associated with band crossings, is constrained by at least one lattice symmetry, and can be turned on/off by breaking or keeping the corresponding lattice symmetry through rotation of the magnetization direction. Local orbital components associated with band crossings need to match the symmetry and produce non-zero spin-orbit coupling when the symmetry is broken. According to this map, TiSb monolayer is predicted to be a quantum anomalous Hall effect material with band gap that can be adjusted in the range of 0 to 209 meV by rotating the magnetization direction.
\end{abstract}

\maketitle

\section{\label{sec:level1} Introduction }
The quantum anomalous Hall effect (QAHE) materials, which break time-reversal symmetry, were first predicted by Haldane in two-dimensional (2D) honeycomb lattices \cite{haldane_model_1988,hasan_colloquium_2010,qi_topological_2011}. The QAHE has attracted widespread attention because of its potential applications in the building of next-generation low-power-consumption electronic devices and the study of novel quantum phenomena such as topological superconductivity and topological magneto-electric effects \cite{liu_quantum_2016,chang_colloquium_2023}. Many pioneering studies have been conducted to experimentally explore and synthesize materials that exhibit QAHE, making a significant contribution to advancing this field \cite{chang_experimental_2013,kou_scale-invariant_2014,chang_high-precision_2015,mogi_magnetic_2015,ou_enhancing_2018}.

The band gap, magnetic properties and topological edge states are three important features of QAHE materials. Many successful attempts have been made to modulate the topological edge state of QAHE materials \cite{zhang_electrically_2012,zhao_tuning_2020,li_chern_2022,zhan_floquet_2023,tang_intrinsic_2023,jiang_large-gap_2023}. It is well known that the band gap is of vital significance in solid state physics and materials science \cite{kittel_introduction_2005}, as it mainly governs the electrical and optical properties of materials. The band gap is also a crucial parameter of QAHE materials relevant to practical applications \cite{wu_prediction_2014,dong_two-dimensional_2016,huang_quantum_2017,li_high-temperature_2020,jiang_monolayer_2024}. A large band gap can make QAHE more stable against disturbances, such as temperature changes or impurity scattering.

The band gap is always modulated by methods that alter the structure and composition, such as strain and doping \cite{conley_bandgap_2013,castellanos-gomez_local_2013}. In contrast to conventional materials, the band gap of QAHE materials originates from the degeneracy of states at band crossings, induced by the spin-orbit coupling (SOC) in materials that break time-reversal symmetry. The size of this band gap is mainly determined by the strength of SOC. It is interesting to adjust the band gap of QAHE material by utilizing its band gap formation mechanism.

Recent reports on spin space groups have greatly enriched the understanding of lattice and spin symmetries \cite{liu_spin-group_2022}. Some interesting phenomena, such as chiral Dirac fermion and altermagnetism \cite{zhang_chiral_2023,smejkal_emerging_2022,feng_anomalous_2022}, are closely associated with these symmetries. The interplay between lattice and spin freedoms introduces innovative ways to manipulate the properties of solids \cite{jiang_spin_2018,liu_intrinsic_2018,li_magnetically_2019,tanttu_controlling_2019,wu_magnetic_2023,sun_tunable_2023,shukla_nodal-line_2023}. Recently, modulating the topological state of QAHE materials through spin or magnetization direction has also attracted attention \cite{zhang_time-reversal_2023}. In this work, we establish the map between symmetries and orbital rules to realize tunable band gap in QAHE material, and TiSb monolayer is predicted to be this type of QAHE material based on the established map.

\section{\label{sec:level1} Computational methods  }

The calculations are performed on the Vienna Ab initio Simulation Package (VASP)\cite{kresse_efficient_1996} within the framework of density functional theory. The projector-augmented wave (PAW) \cite{blochl_projector_1994} is used to describe the interaction between electrons and ions.The generalized gradient approximation (GGA) with Perdew-Burke-Ernzerhof (PBE)\cite{perdew_generalized_1996} is used to describe the interaction between electrons. The spin polarizations of electrons are taken into account in the calculations for the magnetism of transition metals. The effective Hubbard U=3 term is considered in Ti-$d$ orbitals to address the self-interaction error of the generalized gradient approximation \cite{dudarev_electron-energy-loss_1998}. Wave functions of valence electrons are expanded by plane wave basis sets with a cutoff energy of about 400 eV, and the first Brillouin zone is sampled using a $\Gamma$-centered ($11 \times 11 \times 1$) Monkhorst-Pack grid \cite{monkhorst_special_1976}. During structural optimization, lattice parameters and atomic positions are fully relaxed. The convergence standard is set to be about $1 \times 10^{-7}$ eV for total energy and 0.01 eV/\AA\   for the force.

\section{\label{sec:level1} Results and discussion  }

\subsection{\label{sec:level2}Tight-binding Hamiltonian }

\begin{figure}
\includegraphics[width=16 cm]{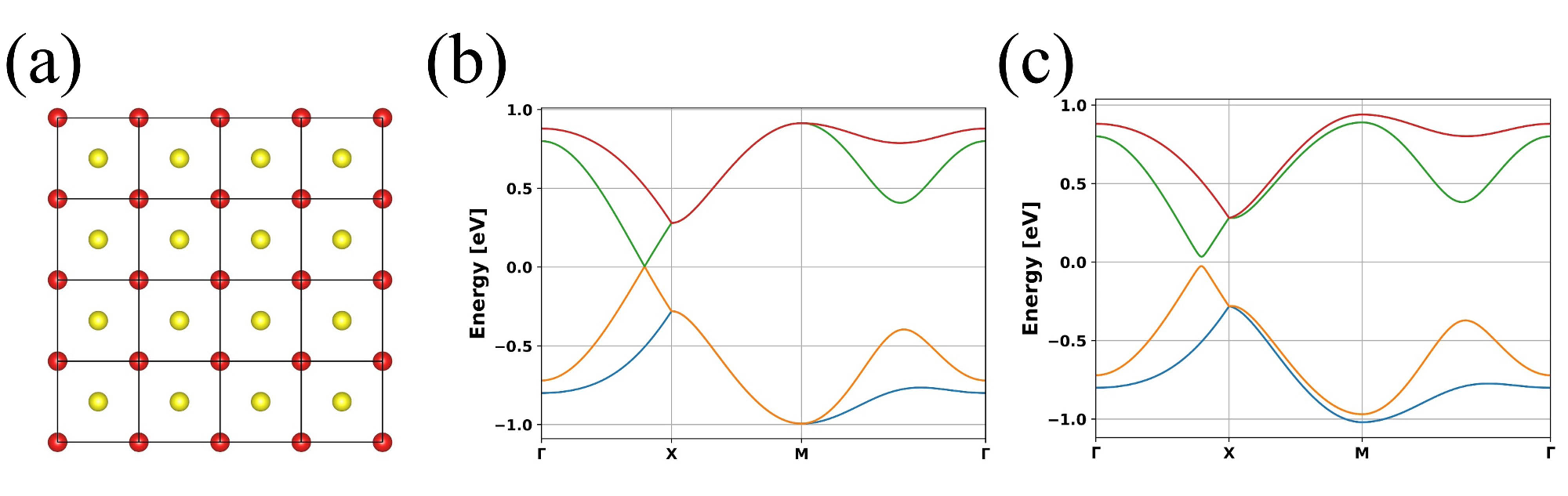}
\caption{\label{tb}  The relationship between band gap and the strength of SOC in a square lattice. (a) A square lattice with two sites at (0.0, 0.0) and (0.5, 0.5). (b) and (c) The band structure without and with SOC ( = 30 meV) based on four orbitals of TB Hamiltonian. The TB parameters are set to be $\varepsilon_1$=-0.28, $\varepsilon_2$=0.28, $t_{11}$(1,0)=0.07, $t_{11}$(0.5,0.5)=0.20, $t_{22}$(1,0)=-0.05, $t_{22}$(0.5,0.5)=-0.20 and $t_{12}$(0.5,0.5)=-0.20 in eV.}
\end{figure}

Band crossing is the origin of many nontrivial topological states \cite{hasan_colloquium_2010,qi_topological_2011,liu_quantum_2016,chang_colloquium_2023}. Considering a 2D ferromagnetic material in a square lattice with two magnetic sites at (0.0, 0.0) and (0.5, 0.5) as shown in Fig.~\ref{tb}(a), there are two local orbitals  $\Phi_1$ and $\Phi_2$ with the same spin at each site. The corresponding tight-binding (TB) Hamiltonian for four bands can be expressed as:

\begin{equation}
\hat H = \sum\limits_{ij,\alpha \beta } {{t_{\alpha \beta }}({r_i} - {r_j})} \hat c_{i\alpha }^ + {\hat c_{j\beta }} + \sum\limits_{i,\alpha \beta } {\left\langle {{\Phi _{i\alpha }}} \right|} {\hat H_{soc}}\left| {{\Phi _{i\beta }}} \right\rangle \hat c_{i\alpha }^ + {\hat c_{i\beta }}
\end{equation}
The first term represents orbital interaction, and the second term represents SOC. While $i$ and $j$ denote sites, $\alpha$ and $\beta$ represent the local orbitals. The Hamiltonian of TB model in momentum space can be formulated as:
\begin{equation}
H(k) = \left( {\begin{array}{*{20}{c}}
{{h_1}}&{{h_{soc}}}&{{h_{11}}}&{{h_{12}}}\\
{h_{soc}^*}&{{h_2}}&{{h_{12}}}&{{h_{22}}}\\
{h_{11}^*}&{h_{12}^*}&{{h_1}}&{{h_{soc}}}\\
{h_{12}^*}&{h_{22}^*}&{h_{soc}^*}&{{h_2}}
\end{array}} \right)
\end{equation}
Where the matrix elements are given by:
\begin{equation}
\left\{ {\begin{array}{*{20}{l}}
{{h_1} = {\varepsilon _1} + 2{{\rm{t}}_{11}}(1,0)[\cos ({k_x}) + \cos ({k_y})]}\\
{{h_2} = {\varepsilon _2} + 2{{\rm{t}}_{22}}(1,0)[\cos ({k_x}) + \cos ({k_y})]}\\
{{h_{11}} = 2{{\rm{t}}_{11}}(0.5,0.5)[1 + {e^{ - i({k_x} + {k_y})}} + {e^{ - i{k_x}}} + {e^{ - i{k_y}}}]}\\
{{h_{22}} = 2{{\rm{t}}_{22}}(0.5,0.5)[1 + {e^{ - i({k_x} + {k_y})}} + {e^{ - i{k_x}}} + {e^{ - i{k_y}}}]}\\
{{h_{12}} = 2{{\rm{t}}_{12}}(0.5,0.5)[1 + {e^{ - i({k_x} + {k_y})}} - {e^{ - i{k_x}}} - {e^{ - i{k_y}}}]}
\end{array}} \right.
\end{equation}
The strength of SOC between local orbitals is defined as:
\begin{equation}
h_{soc} = \left\langle {{\Phi _{i1}}} \right|{\hat H_{soc}}\left| {{\Phi _{i2}}} \right\rangle
\end{equation}
Fig.~\ref{tb}(b) illustrates the band structure without spin-orbit coupling (SOC) ($h_{soc}$ = 0). It is worth noting that band crossings could occur between $\Phi_1$ and $\Phi_2$ type bands along the high symmetry directions $\Gamma X$ and $\Gamma Y$.
The SOC between states related to the band crossing splits the degenerate states of the band crossing and opens band gap ( Fig.~\ref{tb}(c)).

\subsection{\label{sec:level2}Symmetries and orbital rules for tunable band gap}
\begin{figure}
\includegraphics[width=16 cm]{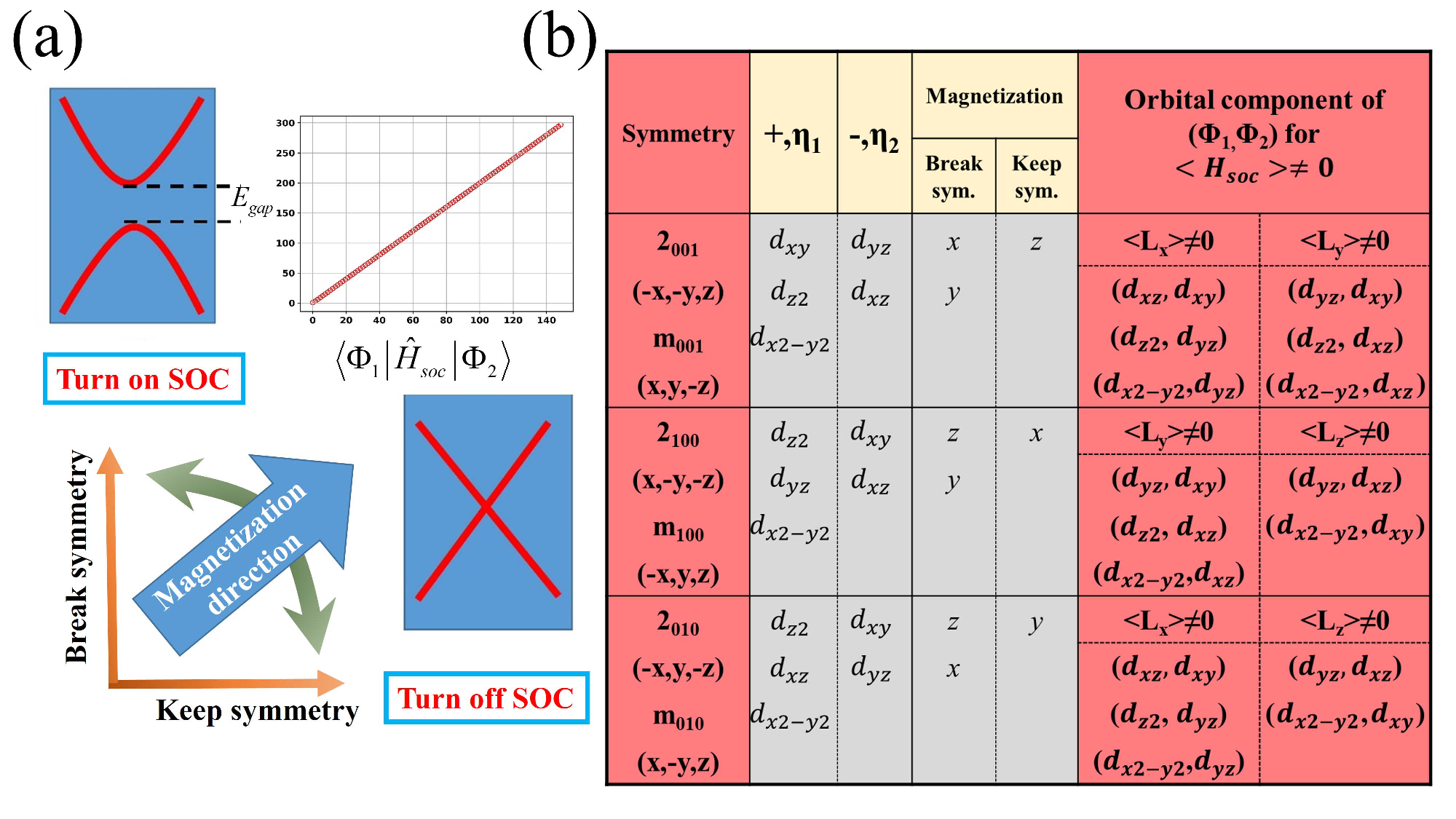}
\caption{\label{map} The map between symmetries and orbital rules to realize a tunable band gap in QAHE materials. (a) The SOC (band gap) is turned on/off by rotating the magnetization direction. (b) The map between symmetries and orbital rules to realize a tunable band gap. $2_{100}$, $2_{010}$ and $2_{001}$ represent $C_2$ rotational symmetries along the $x$, $y$ and $z$ axes, respectively. $m_{100}$, $m_{010}$ and $m_{001}$ indicate the mirror symmetry perpendicular to the $x$, $y$ and $z$  axes, respectively. $\eta _i$ indicates the symmetry of orbital.}
\end{figure}

The local orbitals related to the band crossing have different symmetries, the lattice symmetry imposes a constraint resulting in zero SOC between these local orbitals, as described by the following relationship:
\begin{equation}
\left\langle {{\Phi _1}} \right|{\hat H_{soc}}\left| {{\Phi _2}} \right\rangle  = \left\langle {{\Phi _1}} \right|{\hat P^ + }\hat P{\hat H_{soc}}{\hat P^ + }\hat P\left| {{\Phi _2}} \right\rangle  = \eta _1^*{\eta _2}\left\langle {{\Phi _1}} \right|{\hat H_{soc}}\left| {{\Phi _2}} \right\rangle
\end{equation}
Where $\eta _1$ and $\eta _2$ represent eigenvalues of two local orbitals under symmetry operator $\hat P$ . Thus, the SOC (band gap) of QAHE material could be turned on/off by breaking or preserving this symmetry through the rotation of magnetization direction, as depicted in Fig.~\ref{map}(a). Furthermore, specific local orbitals associated with band crossings are also required to generate non-zero SOC when the symmetry is broken.

We establish the map between symmetries and orbital rules to control the SOC between localized $d$-orbitals of band crossing in QAHE material (Fig.~\ref{map}(b)). Our research focuses on the rotational symmetry $C_2$ and the mirror symmetry $m$, both of which can be completely broken by changing the direction of magnetization. Since the magnetism of materials predominantly originates from $d$ orbitals, our focus remains on these orbitals.

Taking $2_{001}$ or $m_{001}$ as an example, the $z$-direction ($x$- and $y$-directions) magnetization preserves (breaks) the symmetry. If the main components of the orbital pair corresponding to the band crossing are ($d_{xz}$, $d_{xy}$) , ($d_{z^2}$, $d_{yz}$) or ($d_{x^2-y^2}$, $d_{yz}$), the band gap and SOC can be adjusted by rotating the magnetization from $z$ to $x$ direction to break symmetry. If the main components of the orbital pair corresponding to the band crossing are ($d_{yz}$, $d_{xy}$), ($d_{z^2}$,$d_{xz}$) or ($d_{x^2-y^2}$, $d_{xz}$), the tunable band gap can be achieved by rotating the magnetization direction in the $yz$ plane. There are similar requirements for the orbital components for rotational and mirror symmetries along the $x$ and $y$ directions (Fig.~\ref{map}(b)).

From another perspective, due to complete spin polarization, the wave function of the associated state can be expressed as the product of its spatial  $\varphi$ and spin components $\chi $. The strength of SOC between orbitals can also be quantified as the product of orbital momentum $\left\langle {{\varphi _1}} \right|\mathord{\buildrel{\lower3pt\hbox{$\scriptscriptstyle\rightharpoonup$}}
\over L} \left| {{\varphi _2}} \right\rangle $ and spin momentum $\left\langle {{\chi _1}} \right|\mathord{\buildrel{\lower3pt\hbox{$\scriptscriptstyle\rightharpoonup$}}
\over S} \left| {{\chi _2}} \right\rangle $, shown in the following formula:
\begin{equation}
\left\langle {{\Phi _1}} \right|{H_{soc}}\left| {{\Phi _2}} \right\rangle  = {\lambda _{soc}}\left\langle {{\varphi _1}} \right|\mathord{\buildrel{\lower3pt\hbox{$\scriptscriptstyle\rightharpoonup$}}
\over L} \left| {{\varphi _2}} \right\rangle  \cdot \left\langle {{\chi _1}} \right|\mathord{\buildrel{\lower3pt\hbox{$\scriptscriptstyle\rightharpoonup$}}
\over S} \left| {{\chi _2}} \right\rangle .
\end{equation}
The parallelism and perpendicularity of these two momentums correspond to the symmetry-preserving and breaking cases, respectively.

\subsection{\label{sec:level2}The design of QAHE material with tunable band gap}

The map between symmetries and orbital rules offers new insight for discovering or designing QAHE materials with tunable band gaps. To obtain this type of material, we consider a 2D lattice whose point group contains the $C_2$ rotation and mirror symmetry. The challenge is to identify the local orbital that satisfies the map to form band crossing in a stable ferromagnetic monolayer \cite{wu_prediction_2014,dong_two-dimensional_2016,huang_quantum_2017,li_high-temperature_2020,jiang_monolayer_2024}.

\begin{figure}
\includegraphics[width=16 cm]{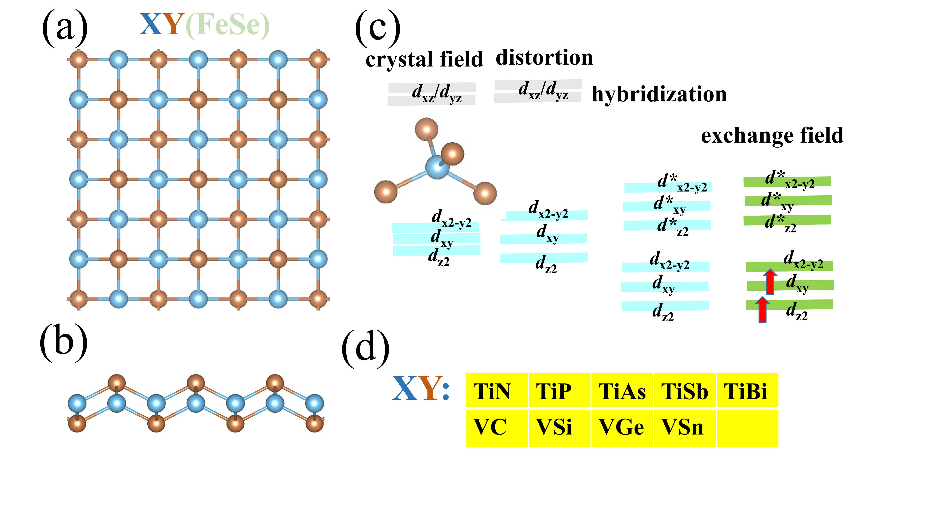}
\caption{\label{design} The design of QAHE material with tunable band gap based on the configuration of FeSe monolayer. (a) and (b) Top and side views of the XY monolayer, respectively. The light blue and golden balls represent X and Y atoms, respectively. (c) The effects of crystal field, hybridization and exchange field on the $d$ orbitals of XY monolayer. (d) The possible components of XY monolayer containing $3d$ transition metal with $d^1$ configuration.}
\end{figure}

We examine the reported monolayers of binary compounds and then utilize the FeSe monolayer \cite{huang_monolayer_2017} as a starting point. The atomic arrangement of the FeSe monolayer meets the symmetry requirements, is shown in Fig. ~\ref{design}(a) and (b). The point symmetries of the paramagnetic, $x$- and $z$-direction magnetization of the FeSe monolayer are presented in Table.\ref{table1}. Rotating the magnetization direction in the $xz$ plane can turn on or off the symmetries of $2_{001}$ and $m_{100}$. The FeSe monolayer is classified in space group No. 129, with each unit cell containing two chemical formulas. Fe (Se) atoms are located in the Wyckoff sites with $-4m2$ ($4mm$) symmetry.

According to the map, we replace Fe (Se) with a 3$d$ transition metal element (main group element) to form an XY monolayer. The X atom is surrounded by four Y atoms, forming a distorted tetrahedron. The effects of the crystal field, hybridization and exchange field on the X-$d$ orbitals, are displayed in Fig. ~\ref{design}(c). Under the effects of the distorted crystal field, the X-$d$ orbitals split into a 2-fold degenerate $d_{xz/yz}$ orbital, three non-degenerate $d_{z^2}$, $d_{xy}$ and $d_{x^2-y^2}$ orbitals. The energies of $d_{z^2}$ and $d_{xy}$ type states are always lower than that of other $d$-type states close to the negatively charged Y atom.

There are two X (Y) atoms in each primitive cell of the XY monolayer. The $d$ orbitals are further hybridized, producing bonding and anti-bonding states. Therefore, if the spin splitting is strong enough and the crystal field effect ($p-d$ hybridization) is greater than the $d-d$ hybridization effect, then the band crossing between $d_{xy} $ and $d_{x^2-y^2}$ type bands is expected in the monolayer containing $3d$ transition metal with $d^1$ configuration. According to the occupation of the $d$ orbital, 9 binary compounds meet these conditions (Fig. ~\ref{design}(d)). By analyzing the band structures of these ferromagnetic configurations, we found that the TiSb monolayer is a candidate for QAHE material that meets the requirements.

\begin{table}
\caption{\label{table1} The point symmetries of the paramagnetic, $x$- and $z$-direction magnetization of FeSe monolayer.}
\begin{ruledtabular}
\begin{tabular}{lcccccccc}
Symmetry& $E$ & $4_{001}$ & $\bar 4_{001}$ & $2_{001}$ & $m_{010}$ & $m_{100}$ & $m_{110}$ &$m_{1{\bar 1}0}$\\
\hline
Paramag. & $\checkmark $ &  $\checkmark$ & $\checkmark$ & $\checkmark$ &  $\checkmark$ & $\checkmark$ & $\checkmark$ &  $\checkmark$ \\
$x$-magnetization &$\checkmark$ & $\times$  & $\times$ & $\times$ & $\times$ & $\checkmark$& $\times$ & $\times$\\
$z$-magnetization & $\checkmark$&  $\checkmark$ & $\checkmark$ & $\checkmark$ &  $\times$ & $\times$ & $\times$ &  $\times$\\
\end{tabular}
\end{ruledtabular}
\end{table}

\subsection{\label{sec:level2} The stability of TiSb monolayer}
\begin{figure}
\includegraphics[width=16 cm]{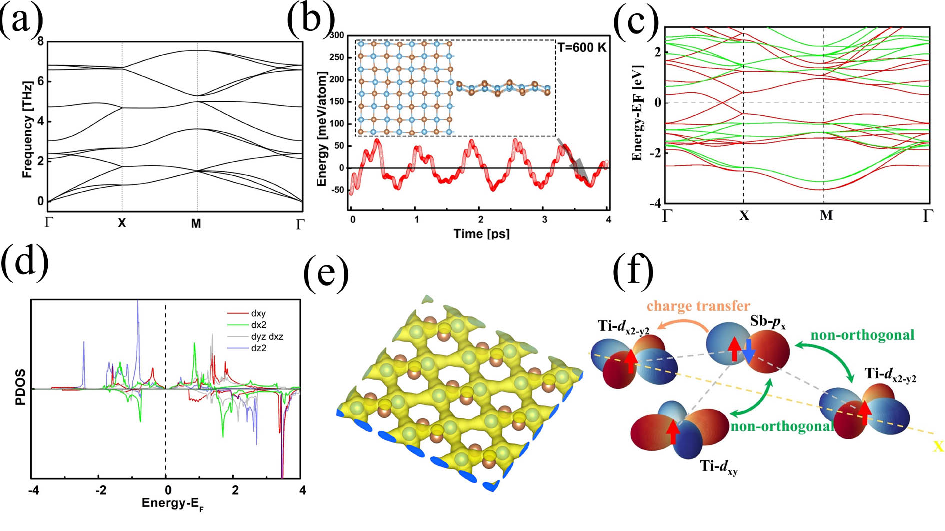}
\caption{\label{eint} The stability, electronic structure and ferromagnetic interactions of the TiSb monolayer. (a) Phonon dispersion of the TiSb monolayer along the high symmetry axis. (b) Total energy obtained from the AIMD simulation performed in a period of 4 ps with NVT ensemble at T = 600 K. (c) Band structure of ferromagnetic TiSb monolayer without SOC. The red and green lines represent the majority and minority channels, respectively. (d) The projected density of states on Ti-$d$ orbitals. (e) The charge distribution of the highest occupied band. (f) Ferromagnetic super-exchange interactions in TiSb monolayer. }
\end{figure}

The optimized in-plane lattice constant and thickness of the TiSb monolayer are determined to be approximately 4.96 and 2.53 \AA, respectively. According to PBE calculations, the formation energy of the TiSb monolayer is approximately -3.72 eV/atom, indicating that the TiSb monolayer is more stable than bulk Sb and Ti. The dynamical and thermal stability of the TiSb monolayer are further assessed through the phonon dispersion (Fig.~\ref{eint}(a)) and ab initio molecular dynamics (AIMD) simulations in the canonical ensemble (NVT) at a temperature of 600 K (Fig.~\ref{eint}(b)), respectively. The TiSb monolayer is located at the minimum of potential energy surface and exhibits no imaginary frequency. Throughout the molecular dynamics simulation, the total energy fluctuations remain below 65 meV/atom, confirming the stability of the TiSb monolayer above room temperature. This suggests that the TiSb monolayer may be obtained by depositing Ti and Sb atoms on a suitable substrate using molecular beam epitaxy (MBE) or pulsed laser deposition (PLD) \cite{cho_molecular_1975,yang_progress_2016}.

\subsection{\label{sec:level2} Electronic interactions and magnetic properties of TiSb monolayer}

The band structure and projected density of states (PDOS) of the ferromagnetic TiSb monolayer without SOC are illustrated in Fig.~\ref{eint}(c) and (d), respectively. Due to the effect of the exchange field, Ti-$d$ states in the minority spin channel are shifted to energy levels above 0.56 eV (Fig.~\ref{eint}(d)). Consequently, the states around the Fermi energy are fully spin-polarized and mainly consist of the majority spin channel of Ti-$d$ states. At the $\Gamma $ point, the bonding states of $d_{z^2}$ and $d_{xy}$ are fully occupied, while the anti-bonding state of $d_{z^2}$ is unoccupied.

\begin{figure}
\includegraphics[width=16 cm]{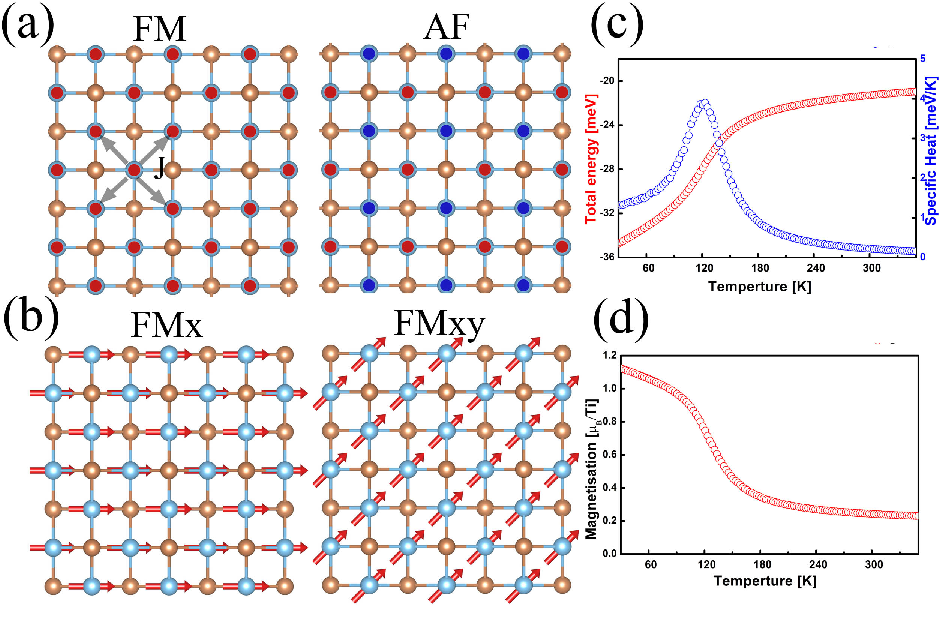}
\caption{\label{mag} The Curie temperature of TiSb monolayer. (a) FM and AF configurations. (b) In-plane FM configurations. (c) and (d), Total energy, specific heat and magnetization of TiSb monolayer as a function of temperature obtained by MC simulation. }
\end{figure}

The charge distribution of the highest occupied band confirms the hybridization between Ti-$d$ and Sb-$p$ orbitals (Fig.~\ref{eint}(e)). According to the Goodenough-Kanamori-Anderson (GKA) rules \cite{goodenough_theory_1955,kanamori_superexchange_1959}, the stability of the ferromagnetic configuration could be attributed to the super-exchange interactions between $d$ orbitals ($d_{xy}$ and $d_{x^2-y^2}$) via the Sb-$p$ orbitals, as presented in Fig.~\ref{eint}(f). This is further supported by the negative magnetic moment of Sb (~-0.15 $\mu_B$) compared to the positive magnetic moment of Ti (~1.17 $\mu_B$). The total energy of the FM configuration is approximately 68 meV/Ti lower than that of the AF configuration (Fig.~\ref{mag}(a)), which is comparable to the case of MnP \cite{wang_mnx_2019}.

The magnetocrystalline anisotropic energy indicates that the ground magnetization of the TiSb monolayer is oriented out-of-plane, which is more energy favorite than the in-plane FM configurations (Fig.~\ref{mag}(b)). The total energy associated with out-of-plane magnetization is about 2.1 meV/Ti lower than that of in-plane magnetization. This is because the orbital moment of Ti for out-of-plane magnetization (~0.22 $\mu_B$) is almost twice that for in-plane magnetization (~0.13 $\mu_B$). Based on Monte Carlo simulation(Fig.~\ref{mag}(c)and (d)), the Curie temperature of TiSb is estimated to be about 123 K, exceeding that of many experimentally studied 2D materials, such as CrI$_3$ (45 K) \cite{huang_layer-dependent_2017} and Cr$_2$Ge$_2$Te$_6$ bilayer (30 K) \cite{gong_discovery_2017}. Epitaxial film growth on a substrate always introduces strain due to lattice mismatch. Strain changes the Curie temperature of TiSb, which can reach room temperature (~297 K) at an in-plane lattice of about 5.1 \AA.

\subsection{\label{sec:level2} Topological properties of TiSb monolayer}
\begin{figure}
\includegraphics[width=16 cm]{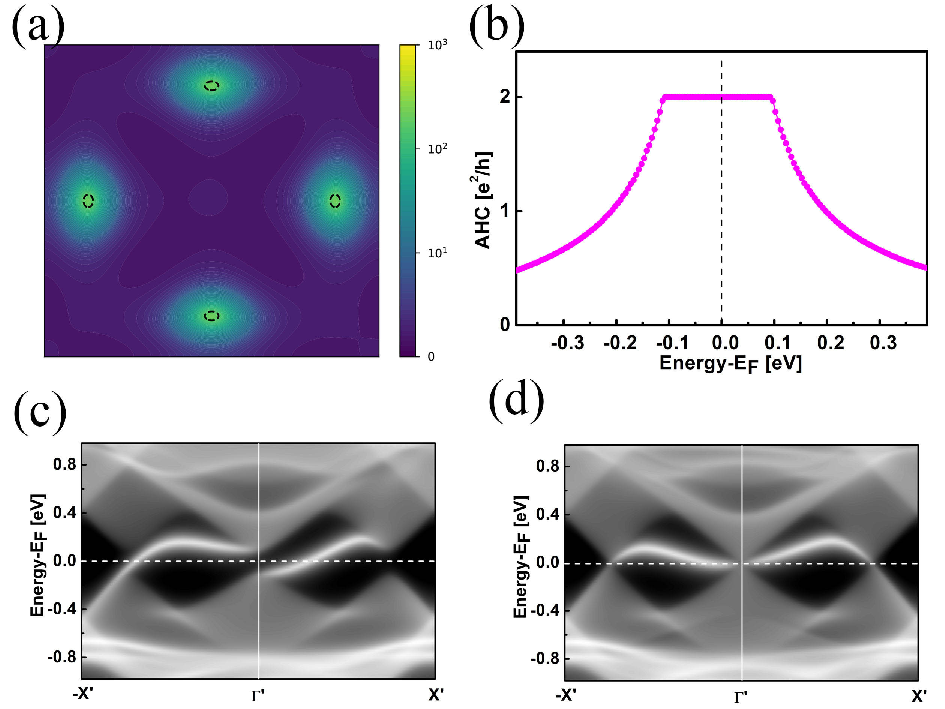}
\caption{\label{QAH} The berry curvature, anomalous Hall conductivity and edge states of TiSb monolayer. (a) Distribution of Berry curvature in momentum space. The isosurface of band energy at -100 meV is shown in dash lines. (b) Anomalous Hall conductivity of the TiSb monolayer as a function of energy around the Fermi energy. (c) and (d) The edge states along the $x$ direction in the $FM_z$ and $FM_x$ configurations, respectively. }
\end{figure}

To verify the topological non-triviality of the TiSb monolayer, the distribution of Berry curvature in the 2D Brillouin zone is presented in Fig.~\ref{QAH}(a). Throughout the Brillouin zone, Berry curvature has four extreme points near the band crossings along the $\Gamma X$ and $\Gamma Y$ axes. The intrinsic anomalous Hall conductivity is quantized (${\sigma _{xy}} = \frac{{2{e^2}}}{h}$ ) when the Fermi level lies inside the bulk band gap, and is close to zero when the Fermi level is outside the gap, as shown in Fig. ~\ref{QAH}(b).

The topological edge states of the out-of-plane magnetized TiSb monolayer along the $x$ direction are calculated using Green's functions based on the Wannier function\cite{mostofi_updated_2014} to reduce the computational cost, as presented in Fig.~\ref{QAH}(c). The band structure below the Fermi energy forms 4 cones in the Brillouin zone. When projected to the edge in the $x$ direction, the 2 cones along the $\Gamma Y$ direction merge. Thus, there are 3 cones below Fermi energy in the projected band structure of 100 edge. There are two edge states in each edge connecting the valence and conduction bands, which is in good agreement with the results of intrinsic anomalous Hall conductivity. In addition, we also calculate the edge states of in-plane magnetization, as shown in Fig.~\ref{QAH}(d), which confirms the non-trivial topological properties of the corresponding states.

\subsection{\label{sec:level2} Tunable band gap of TiSb monolayer}

\begin{figure}
\includegraphics[width=16 cm]{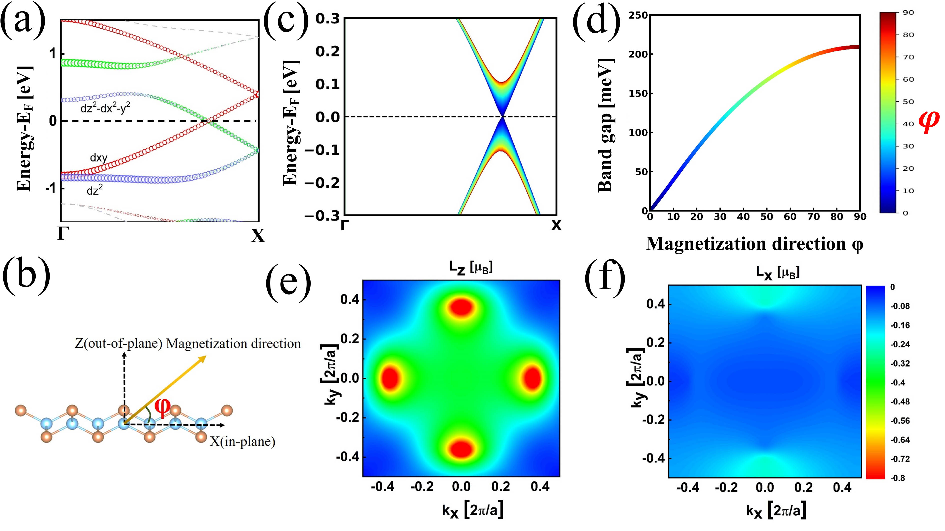}
\caption{\label{tgap} Tunable band gap in TiSb monolayer. (a) Orbital-resolved bands along $\Gamma X$ without SOC. (b) Schematic representation of magnetization direction. (c) Evolution of bands along $\Gamma X$  with the rotation of magnetization-direction ($\varphi $). (d) The band gap as a function of magnetization direction. (e) and (f) Orbital momentum distribution of the highest occupied band in the BZ with the $z$- and $x$-direction magnetization, respectively. }
\end{figure}

As shown in Fig.~\ref{tgap}(a), the energy of the $d_{xy}$ type band is about 1.11 eV lower (0.83 eV higher) than that of the $d_{z^2}$-$d_{x^2-y^2}$ type band at the $\Gamma (X)$ point. The band crossing along the  $\Gamma X$ direction is formed between the $d_{xy}$ and $d_{z^2}$-$d_{x^2-y^2}$ type bands, giving rise to a non-trivial topological state. It is worth noting that both states at the band crossing belong to different irreducible representations of the Little $K$ group. This strictly defines band crossing without considering SOC. This band crossing is also confirmed by HSE calculations \cite{heyd_hybrid_2003,heyd_efficient_2004}.

The TiSb monolayer exhibits $m_{100}$ symmetry without consideration of magnetism. According to the symmetries and orbital rules (Fig.~\ref{map}(b)), as expected, the SOC between the corresponding local orbitals could be turned on/off by rotating the magnetization direction in the $xz$ plane. The band crossing of the TiSb monolayer remains in the $x$ magnetization, and opens a band gap of about 209 meV in the $z$ magnetization (Fig.~\ref{tgap}(b)$\sim$(d)). The band gap gradually decreases to zero as the angle of magnetization direction $\varphi$ decreases.

The magnetization direction also affects orbital momentum. We calculate the orbital momentum distribution of the highest occupied state in the Brillouin zone (BZ) with $z$- and $x$-direction magnetization (Fig.~\ref{tgap}(e) and (f)), using the following formula:
\begin{equation}
\left\langle {{\rm{\mathord{\buildrel{\lower3pt\hbox{$\scriptscriptstyle\rightharpoonup$}}
\over L} }}} \right\rangle  = \sum\limits_{k,i,lm,l'm'} {{f_{nk}}\left\langle {nk} \right|\left. {ilm} \right\rangle \left\langle {ilm} \right|{\rm{\mathord{\buildrel{\lower3pt\hbox{$\scriptscriptstyle\rightharpoonup$}}
\over L} }}\left| {il'm'} \right\rangle \left\langle {il'm'} \right|\left. {nk} \right\rangle } .
\end{equation}
Where $f_{nk}$ is the occupation number of the state, while $\left| {nk} \right\rangle$ and  $\left| {ilm} \right\rangle$ represent the wave functions and local orbits, respectively. In the case of $z$-direction magnetization, the orbital momentum mainly concentrates around band crossings. In contrast, with $x$-direction magnetization, the orbital momentum decreases and is confined close to zero near the $k_x$ axis due to the symmetry of $m_{100}$. These findings provide a qualitative analysis \cite{thonhauser_orbital_2005,shi_quantum_2007,lopez_wannier-based_2012}, suggesting that the magnetization direction influences the orbital momentum and alters the strength of SOC. It can be understood that the crystal field may dampen the orbital momentum of the central ion, and the magnetization direction can break the symmetry of the crystal field, thereby restoring or partially restoring the orbital momentum.

\section{\label{sec:level1} Conclusion  }

In summary, we establish the map between symmetries and orbital rules to realize tunable band gaps in QAHE materials. Starting from the band crossing, the local orbitals of the band crossing belong to different irreducible representations of the lattice symmetry group, and have different eigenvalues under at least one lattice symmetry operation. The band gap is determined by the SOC between local orbitals associated with band crossing, and is constrained by the corresponding lattice symmetries. Thus, the SOC (band gap) of QAHE materials could be turned on/off by the rotation of the magnetization direction to maintain or break the corresponding symmetry. In addition to this, the local orbital components associated with band crossings need to match the symmetry and produce non-zero SOC when the symmetry is broken.

Following the map, the TiSb monolayer is predicted to be a QAHE material with an adjustable band gap. The stability of the TiSb monolayer is confirmed by the results of formation energy, phonon dispersion and AIMD simulation. The band gap of the TiSb monolayer can be tuned from 0 to 209 meV by breaking $m_{010}$ symmetry through the rotation of magnetization direction. Our findings not only provide novel insights into the design of QAHE materials with tunable band gaps, but also open avenues for future exploration of materials capable of achieving broader tunable band gaps.

\begin{acknowledgments}
We thank Li Ren and Huiping Wang for helpful discussions. This work is supported by The National Natural Science Foundation of China (NSFC, NO.12174247).
\end{acknowledgments}

\section*{Author Declarations}
The authors have no conflicts to disclose.

\section*{Data Availability Statement}
The data that support the findings of this study are available from the corresponding authors upon reasonable request.

\nocite{*}
\bibliography{TiSb}

\end{document}